\documentclass{article}
\pdfoutput=1

\usepackage[utf8]{inputenc}
\usepackage{amssymb}
\usepackage{hyperref,braket}
\usepackage{amsmath, amsthm, graphicx}
  {

  }
  \usepackage{color}
\usepackage[sorting=none,giveninits=true]{biblatex}
\theoremstyle{definition}

\theoremstyle{definition}

\def\s[#1\s]{\begin{align}\begin{split}#1\end{split}\end{align}}
\usepackage{xspace}

\DeclareMathOperator{\e}{e}

\allowdisplaybreaks
\graphicspath{ {images/} }

\begin{document}

\begin{titlepage}
\title{\hfill\parbox{4cm}{ \normalsize YITP-22-114}\\ 
\vspace{1cm} 
The tensor of the exact circle: Reconstructing geometry.}
\author{Dennis Obster\footnote{dennis.obster@yukawa.kyoto-u.co.jp} \\
{\small{\it Yukawa Institute for Theoretical Physics, Kyoto University,}}
\\ {\small{\it  Kitashirakawa, Sakyo-ku, Kyoto 606-8502, Japan}}
}

\date{November 2022}

\maketitle

\begin{abstract}
    Developing a theory for quantum gravity is one of the big open questions in theoretical high-energy physics. Recently, a tensor model approach has been considered that treats tensors as the generators of commutative non-associative algebras, which might be an appropriate interpretation of the canonical tensor model. In this approach, the non-associative algebra is assumed to be a low-energy description of the so-called associative closure, which gives the full description of spacetime including the high-energy modes. In the previous work it has been shown how to (re)construct a topological space with a measure on it, and one of the prominent examples that was used to develop the framework was the exact circle. In this work we will further investigate this example, and show that it is possible to reconstruct the full Riemannian geometry by reconstructing the metric tensor. Furthermore, it is demonstrated how diffeomorphisms behave in this formalism, firstly by considering a specific class of diffeomorphisms of the circle, namely the ellipses, and subsequently by performing an explicit diffeomorphism to ``smoothen'' sets of points generated by the tensor rank decomposition.
\end{abstract}

\end{titlepage}

\section{Introduction}
In the standard model of particle physics, three of the four fundamental forces of nature are described with high accuracy. The model is the currently best fundamental understanding of the interaction of particles, and fits experimental results extremely well, for example with the discovery of the Higgs-boson~\cite{ATLAS:2012yve,CMS:2012qbp}. Notably, gravity is absent in this fundamental small-scale understanding of the universe. On the other hand, general relativity is the best understanding of gravity currently available, and on macroscopic scales it describes the dynamics of nature up to very high accuracy~\cite{EinsteinMercury1915,Will2014,Dyson:1920cwa, Abbott:2016blz}. However, one of the current big open questions in theoretical physics is how to combine these two theories to fully understand gravity and the interaction of particles.

Ultimately, the reason for gravity to not be included in this picture yet is that general relativity is perturbatively non-renormalisable~\cite{tHooft:1974toh,Goroff:1985th}. Many approaches have been tried to solve these issues~\cite{Loll:2022ibq}, including reformulating renormalisation using the exact renormalisation group~\cite{Weinberg:1980gg, Reuter_PhysRevD.57.971, Reuter:2012id}, reformulating gravity in terms of Ashtekar variables~\cite{rovelli1988new, Thiemann}, and discrete approaches~\cite{Ambjorn:1998xu, Ambjorn:2012jv, Loll:2019rdj,sorkin1991spacetime}. The original tensor models~\cite{Sasakura:1990fs,Ambjorn:1990ge, Godfrey:1990dt} are an example of a discrete approach, where a spacetime is constructed by gluing simplices together according to the contraction of tensors. These models have not been very successful as a theory for quantum gravity yet, with the absense of emerging macroscopic spacetimes, though there has been some interesting advancements of the so-called coloured tensor model approach which might relate to gravity through holography~\cite{Gurau:2009tw}.

Motivated by the success of causal dynamical triangulation, which is a discrete approach where a notion of causality plays a central role, the canonical tensor model was introduced in order to introduce a notion of causality into the framework of tensor models~\cite{Sasakura:2011sq,Sasakura:2012fb,Sasakura:2013gxg,Sasakura:2013wza}. This model has had some very interesting results, finding connections to general relativity~\cite{Sasakura:2014gia, Sasakura:2015pxa, Chen:2016ate}, finding exact wave functions~\cite{Sasakura:2013wza, Narain:2014cya}, and the analysis of these wave functions~\cite{Obster:2017pdq,Obster:2017dhx,Lionni:2019rty,Sasakura:2019hql,Kawano:2021vqc,Obster:2020vfo,Sasakura:2021lub}. The model is, however, difficult to interpret in a similar way as the original tensor models. Therefore, a new approach was introduced in~\cite{Obster:2022oal} to make a direct link to topological spaces. In this approach, a tensor generates an algebra of functions of some topological space, which is supposed to represent a spatial slice of spacetime, making the connection to general relativity through the ADM formalism~\cite{Arnowitt1959}. Using the duality between manifolds and the algebra of smooth functions on them~\cite{nestruev2006smooth}, the manifold may be reconstructed from this tensor. However, to arrive at a full picture of gravity, the metric tensor on a Riemannian manifold needs to be reconstructed. The method is well-suited as an interpretation for tensor models in the Hamiltonian framework such as the canonical tensor model, since only tensors corresponding to Riemannian manifolds have been considered sofar, as opposed to pseudo-Riemannian manifolds. These tensors could correspond to spatial slices of spacetime to complete the full spacetime picture.

In this work, we describe a way to encode the full geometry of a (compact) Riemannian manifold in a tensor $P_{abc}$. The general idea is to encode the spectrum of the Laplace-Beltrami operator in the tensor, and define a procedure to extract it. Due to the intimate connection between the Laplace-Beltrami operator and the metric~\cite{rosenberg_1997}, the full metric may be reconstructed in this way. We will describe the general procedure, and work it out completely for the example of the exact circle. Furthermore, the action of diffeomorphisms is considered and shown to be consistent, by describing a class of spaces diffeomorphic to the circle: Ellipses. Lastly an explicit discrete diffeomorphism is performed to ``smoothen out'' randomly generated sets of points.

This work is structured as follows. In section~\ref{sec:exact_circle}, the geometry of the circle is described, and the algebra of smooth functions over the circle is considered. In section~\ref{sec:tensor_def}, the relevant points of the algebraic tensor models are briefly reviewed, including the inclusion of the spectrum of the Laplace-Beltrami operator. In section~\ref{sec:reconstruct}, we show the reconstruction of the metric using smooth function and the spectrum of the Laplace-Beltrami operator. In section~\ref{sec:diffeo}, this is then applied to the algebraic tensor model framework of~\cite{Obster:2022oal}, and it is shown that knowing a finite-dimensional tensor $P_{abc}$ is enough to reconstruct the full Riemannian circle. Section~\ref{sec:sum} then concludes this work.

\section{The geometry of the exact circle}\label{sec:exact_circle}
In this section we will closely examine the geometry and properties of the exact circle. The circle will be defined, and the metric properties and the smooth functions on the circle will be explained. Lastly the action of the Laplace-Beltrami operator on the circle is considered.

The exact circle is a closed one-dimensional Riemannian manifold $(S^1, g)$. Any (connected) closed one-dimensional manifold is homeomorphic to the circle. We consider the unit-circle as the set of points $(x,y)\in \mathbb{R}^2$, such that
\begin{equation*}
    x^2 + y^2 = 1,
\end{equation*}
with the induced topology and induced metric from $\mathbb{R}^2$. In local coordinates $\theta \in [0,2\pi )$, the metric is given by 
\begin{equation}\label{eq:circ_geom:g}
    g_{\theta \theta } = 1.
\end{equation}
In general in gravitational physics, one considers all diffeomorphisms of the manifold mentioned above.

As might be expected from the topology of the circle, the algebra of real smooth functions on the circle, $C^\infty(S^1)$, is identical to the algebra of $2\pi$-periodic smooth functions on $\mathbb{R}$
\begin{equation*}
    C^\infty(S^1) \cong \{ f \in C^\infty(\mathbb{R}) |\, f(\theta+2 \pi) = f(\theta) \}.
\end{equation*}
The algebra of real smooth functions $C^\infty(S^1)$ is an infinite-dimensional vector space, together with a pointwise product that is given by $f,g\in C^\infty(S^1)$
\begin{equation*}
    (f\cdot g)(\theta) = f(\theta)g(\theta).
\end{equation*}
An example of such a product would be
\begin{equation*}
    \sin(\theta) \sin(\theta) = \frac{1}{2}\left(1-\cos(2\theta)\right)
\end{equation*}
Since the circle is a compact Riemannian manifold, it carries a natural measure, infinitesimally ${\rm d}\theta$, with which we can define an inner product on the algebra of smooth functions
\begin{equation}\label{eq:circ_geom:innerp}
    \braket{f|\, g} := \int_{S^1} {\rm d}\theta f(\theta) g(\theta).
\end{equation}
The closure of the smooth functions $C^\infty(S^1)$ with respect to this inner product gives the square integrable functions $L^2(S^1)$.

On the space of square integrable functions, one can define the Laplace-Beltrami operator\footnote{Note that operators on Hilbert spaces are defined on dense subsets of the Hilbert space~\cite{conway1994course}, in this case the twice-differentiable functions $C^2(S^1)$.}
\begin{equation*}
    \Delta : L^2(S^1) \rightarrow L^2(S^1).
\end{equation*}
In general, this operator is given in local coordinates by
\begin{equation}\label{eq:circ_geom:operator}
    \Delta f := \frac{1}{\sqrt{\det g}} \partial_j \left( g^{ij} \sqrt{\det g} \partial_i f \right),
\end{equation}
which on the exact circle with metric~\eqref{eq:circ_geom:g} simply reduces to
\begin{equation*}
    \Delta f = \partial_\theta^2 f.
\end{equation*}
The Laplace-Beltrami operator is actually self-adjoint with respect to the inner product of~\eqref{eq:circ_geom:innerp}. The eigenfunctions of the operator span a (Schauder) basis $\{f_n\}$ for $L^2(S^1)$, and are given by
\begin{equation}
\{ f_1=\frac{1}{\sqrt{2\pi}},\,  f_2=\frac{1}{\sqrt{\pi}} \sin(\theta),\,  f_3=\frac{1}{\sqrt{\pi}} \cos(\theta),\, f_4=\frac{1}{\sqrt{\pi}} \sin(2\theta),\, f_5=\frac{1}{\sqrt{\pi}} \cos(2\theta),\ldots \},\label{eq:circle_basis}
\end{equation}
with $\theta\in [0,2\pi)$ and the functions were normalised with respect to the inner product~\eqref{eq:circ_geom:innerp}. The eigenvalues, denoted by $\lambda_n$ for $n\geq 1$, of the Laplace-Beltrami operator corresponding to those functions are given by~\cite{zelditch2017eigenfunctions}
\begin{equation}\label{eq:laplace_eigenv}
\begin{aligned}
    \lambda_{2n+1} &= n^2,\\
    \lambda_{2n} &= n^2.
\end{aligned}
\end{equation}

The main question is, if one only has the algebra of smooth functions $\mathcal{A} \cong C^\infty(S^1)$, i.e. only an abstract vector space with a product operation
\begin{equation*}
    \cdot : \mathcal{A} \times \mathcal{A} \rightarrow \mathcal{A},
\end{equation*}
can one reconstruct the full structure of the Riemannian manifold one started with?  One can separate this question into two parts, how to reconstruct the topological structure and how to reconstruct the metric. The topological part can be reconstructed from the algebra and the product alone~\cite{nestruev2006smooth}. This is done in the following way. 

First, one considers the ``set of points'' of the topological space. Assume some topological manifold $\mathcal{M}$, with the smooth functions on this manifold $C^\infty(\mathcal{M})$. The points of the manifold are found by the realisation that every point $x\in\mathcal{M}$ corresponds to an evaluation map $p_x$ on the smooth functions
\begin{equation*}
\begin{aligned}
    p_x : C^{\infty}(\mathcal{M}) &\rightarrow \mathbb{R},\\
    p_x(f) &:= f(x).
\end{aligned}
\end{equation*}
These evaluation maps are naturally $\mathbb{R}$-homomorphisms of the algebra, by definition of the pointwise product, since
\begin{equation*}
    p_x(f\cdot g) = (f\cdot g)(x) = f(x) g(x) = p_x(f)p_x(g).
\end{equation*}
Defining the algebraic dual space, $|\mathcal{A}|$, of an abstract algebra, $\mathcal{A}$, is now done by taking all $\mathbb{R}$-homomorphisms of the algebra
\begin{equation}\label{eq:algebraic_dual}
    |\mathcal{A}| := \left \{ p : \mathcal{A} \rightarrow \mathbb{R} |\, p\text{ is a homomorphism} \right\}.
\end{equation}
$|\mathcal{A}|$ now corresponds to the set of points of the prospective manifold, which we wish to equip with a topology. This is done by taking the weakest topology such that the elements of $\mathcal{A}$ become continuous~\cite{nestruev2006smooth}, which yields a Hausdorff-topology. In this work we will not go further into the full manifold structure, but it can be shown that if the algebra corresponds to the algebra of smooth functions on a manifold, i.e. $\mathcal{A} \cong C^\infty(\mathcal{M})$, then the full manifold structure may be reconstructed.

\begin{figure}
    \centering
    \makebox[\textwidth][c]{\includegraphics[width=1.3\textwidth]{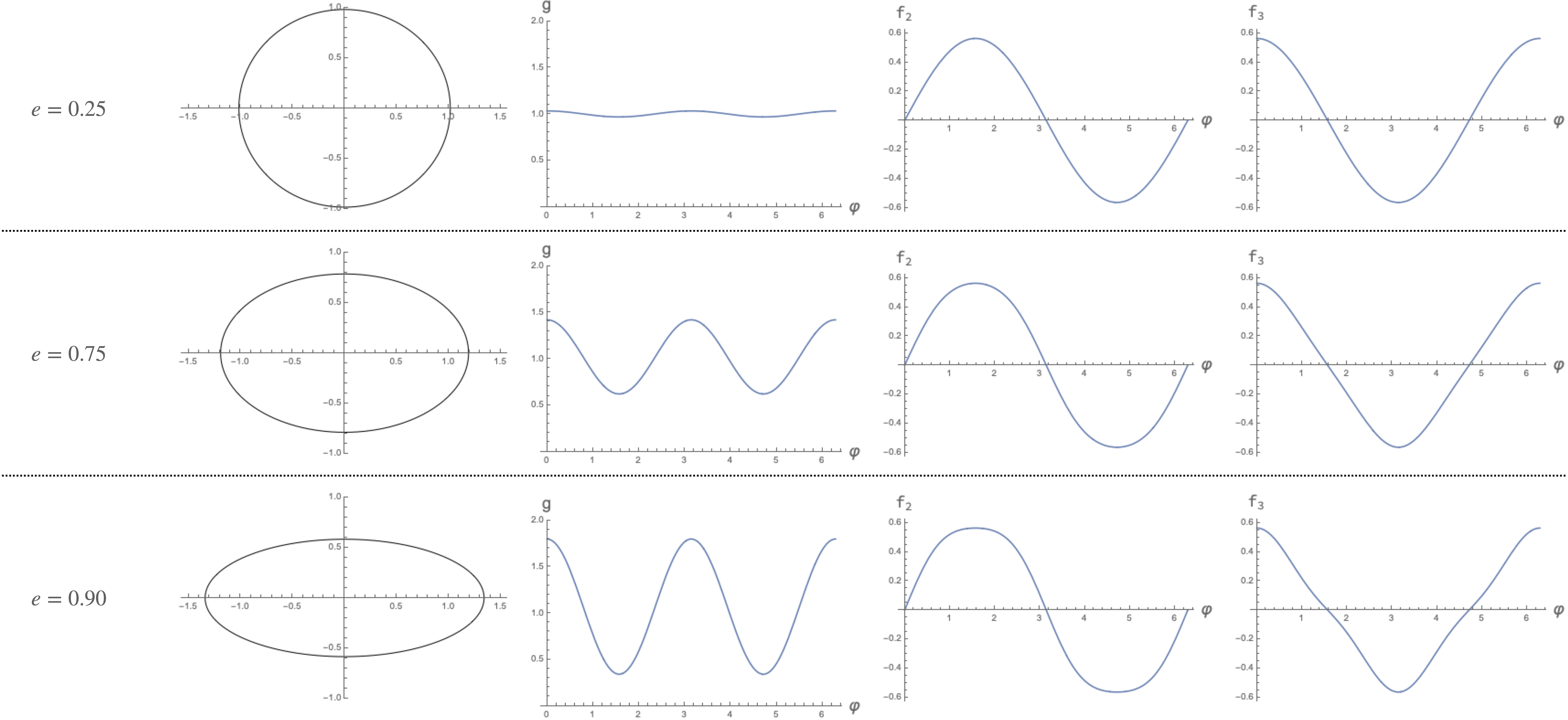}}
    \caption{This figure shows three different ellipses with different values of $e$, as explained in the text. The left image shows the geometrical shape as a subset of $\mathbb{R}^2$. The second image shows the metric as a function of the 1-dimensional variable $\varphi$. The two figures on the right show the functions $f_2(\varphi)$ and $f_3(\varphi)$, which are the first two non-constant eigenfunctions of the Laplace-Beltrami operator.}
    \label{fig:ellipse_example_functions}
\end{figure}

For the geometric part, the information in the algebra $C^\infty(\mathcal{M})$ does not suffice. However, it is known that if one adds spectral data of certain operators on the function space, the full geometry, i.e. the metric, may be reconstructed. Here we will mainly focus on the Laplace-Beltrami operator.

From~\eqref{eq:circ_geom:operator} it is already clear that the Laplace-Beltrami operator contains the information of the metric. Expanding this equation yields
\begin{equation}\label{eq:laplace_approx}
\begin{aligned}
    \Delta f &:= \frac{1}{\sqrt{\det g}} \partial_j \left( g^{ij} \sqrt{\det g}\, \partial_i f \right),\\
    &\approx g^{ij} \partial_i \partial_j f + \text{lower order}.
\end{aligned}
\end{equation}
From this it seems clear that if one takes a function that is locally $f \sim x^i x^j$, one can find the values of $g^{ij}$. This means that not only is the Laplacian dependent on the metric, but knowing the action of the Laplacian on functions also determines the metric. 

As a proof of concept of the reconstruction of the metric in the following sections, we will consider a class of diffeomorphisms of the circle, namely the ellipse. The ellipses are a class of one-dimensional manifolds characterised by a parameter $e \in [0,1)$,\footnote{We are only considering ellipses with the same circumference as the unit circle, otherwise there would be a second parameter.}, where taking $e=0$ reproduces the circle. The ellipse has a half-width of $a(e) = \frac{\pi}{2 E(e^2)}$ and half-height $b(e)=a \sqrt{1-e^2}$. Here, $E(e)$ is the complete elliptic integral of the second kind
\begin{equation*}
    E(e) = \int_{0}^{\frac{\pi}{2}}\sqrt{1-e^2 \sin(\theta)^2} {\rm d} \theta.    
\end{equation*}
The metric of the ellipse is given by
\begin{equation}
    g_{\varphi\varphi}(\varphi) = a(e)^2 (1-e^2 \sin(\varphi)^2).\label{eq:metric_ellipse}
\end{equation}
The functions on the ellipse are similar to the typical $\sin$ and $\cos$ functions on the circle, just deformed. This is because the ellipse may be constructed by a diffeomorphism of the exact circle
\begin{equation}
    \theta(\varphi) = a(e) E(\varphi, e),\label{eq:diffeo_ellipse}
\end{equation}
where $E(\varphi,e)$ is the incomplete elliptic integral of the second kind
\begin{equation*}
    E(\varphi,e) = \int_{0}^{\varphi}\sqrt{1-e^2 \sin(\theta)^2} {\rm d} \theta.    
\end{equation*}
An example of three of these diffeomorphisms is given in figure~\ref{fig:ellipse_example_functions}. It can be seen that for a small $e$, the functions and metric are still very similar to the circle, but for larger $e$ one starts to see larger deformations. Since the transformation is a diffeomorphism that does not change the circumference, the eigenvalues of the ellipse are the same as the circle.

\section{Tensors of the exact circle and ellipses}\label{sec:tensor_def}
In this section, the tensor corresponding to the exact circle as introduced in~\cite{Obster:2022oal} will be discussed. We will introduce two versions of the tensor, with and without the spectral data of the Laplace-Beltrami operator, and it will be shown how diffeomorphisms are to be treated.

Consider the exact circle as discussed in section~\ref{sec:exact_circle}, with the basis elements $\{f_a\}_{a\geq 1}$ as in~\eqref{eq:circle_basis}. In~\cite{Obster:2022oal}, the tensor of the exact circle, $P_{abc}$, acting on a $N$-dimensional real vectorspace $\mathcal{F}\cong \mathbb{R}^N$, was defined as
\begin{equation*}
    P_{abc} := \braket{f_c |\, f_a \cdot f_b},
\end{equation*}
for $a,b,c \leq N$. Here the $N=5$-dimensional version is considered, as it is the smallest non-trivial example. In~\cite{Obster:2022oal} it was argued that by taking the ``associative closure'' of the tensor $P_{abc}$, one arrives at an infinite-dimensional algebra $\mathcal{A} \cong C^{\infty}(\mathcal{S}^1)$. For a brief discussion on the definition of the associative closure, see appendix section~\ref{sec:app}. With this new algebra $\mathcal{A}$, the tensor $P_{abc}$ corresponds to the structure constants of the algebra when restricted to $\mathcal{F}$
\begin{equation*}
    f_a \cdot f_b = \sum_{c\geq 1} P_{abc} f_c.
\end{equation*}

The unit of the tensor $P_{abc}$ is defined as the element $1=\alpha^a f_a \in\mathcal{F}$, such that
\begin{equation*}
    \delta_{ab} = \alpha^c P_{abc}.\label{eq:def_unit}
\end{equation*}
Note that it is not guaranteed that such a unit exists, which will be important for the inclusion of the spectral data below. For now, the tensor is assumed to be such that there is such a unit.

As was conjectured in~\cite{Obster:2022oal}, the associative closure may conveniently be found - if the tensor corresponds to a Riemannian manifold - by using a tool from data analysis called the tensor rank decomposition that has been used in the analysis of the canonical tensor model numerous times~\cite{Sasakura:2021lub,Kawano:2021vqc,Kawano:2018pip,Obster:2021xtb,Sasakura:2022aru}, which for a given tensor $P_{abc}$ is a decomposition of this tensor into $R$ rank-1 tensors
\begin{equation*}
    P_{abc} = \sum_{i=1}^R \beta_i p^i_a p^i_b p^i_c,
\end{equation*}
where $p^i\in \mathbb{R}^N$. This decomposition is called minimal if the number $R$ is the smallest $R$ possible to find such a decomposition, and positive if $\beta_i>0$ for all $\beta_i$. In this case, the decompositions have to be required to be minimal and to be consistent with the inner product in the sense that
\begin{equation*}
    \braket{f_a|\, f_b} = \delta_{ab} = \sum_{i=1}^R \beta_i p^i_a p^i_b,
\end{equation*}
and have $\beta_i>0\, \forall\, 1\leq i \leq R$, such that the $\beta_i$ may be interpreted as a measure. The elements $p_a^i$ are interpreted as candidate points, and are called \emph{potential homomorphisms} (see appendix section~\ref{sec:app}). In the associative closure, these potential homomorphisms will correspond to points through the Gelfand transform, where points are seen as the evaluation maps
\begin{equation*}
    p_a^i \equiv f_a(p^i).
\end{equation*}
The tensor rank decomposition is not unique, and every tensor rank decomposition will give a new set $p_a^i$ that are part of the algebraic dual space of the closure algebra when restricted to $\mathcal{F}$. The elements $p_a^i$ may thus be interpreted as the points of the manifold, and by adding new elements to the vectorspace in order to close the pointwise algebra one arrives at the associative closure.

\begin{figure}
    \centering
    \makebox[\textwidth]{\makebox[1.32\textwidth]{\begin{minipage}{0.4\textwidth}
            \includegraphics[width=0.95\textwidth]{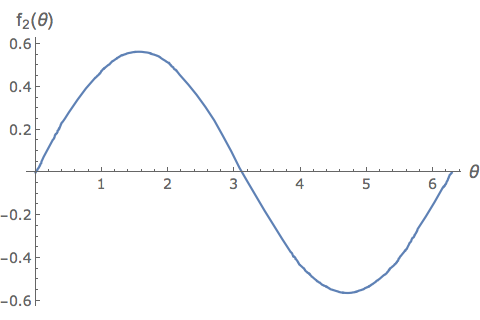}
        \end{minipage}
        \begin{minipage}{0.4\textwidth}
            \includegraphics[width=0.95\textwidth]{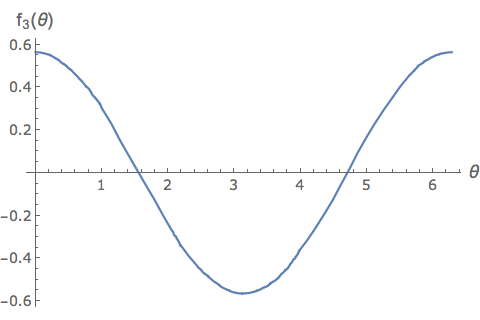}
        \end{minipage}
        \begin{minipage}{0.5\textwidth}
            \includegraphics[width=0.95\textwidth]{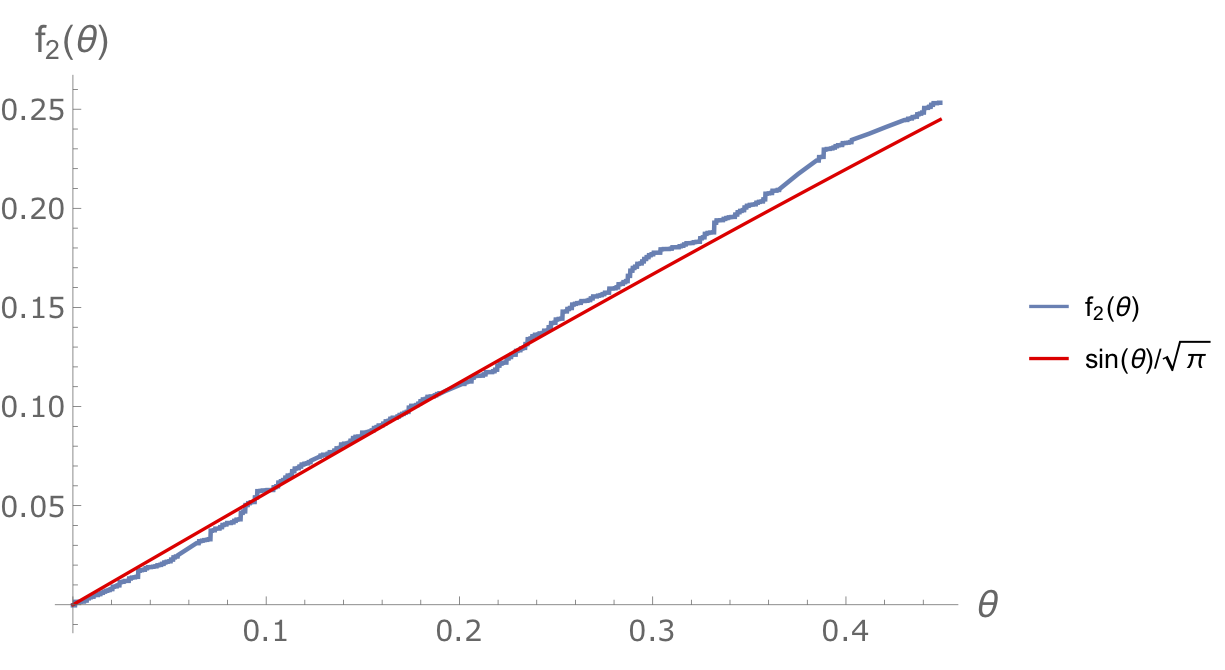}
        \end{minipage}}}
    \caption{Plots of 5208 potential homomorphisms of the tensor of the exact circle, taken from 744 tensor rank decompositions. On the left and middle, the simple function approximation of $f_2(\theta)$ and $f_3(\theta)$ is given. They indeed closely resemble the functions they are supposed to represent, $\sin(\theta)/\sqrt{\pi}$ and $\cos(\theta)/\sqrt{\pi}$ respectively. The first point is chosen such that the phase matches these functions. On the right, the function is zoomed in on to show the deviation of the simple function from $\sin(\theta)/\sqrt{\pi}$.}
    \label{fig:TRDpothomo}
\end{figure}
By taking a finite amount of tensor rank decompositions, one effectively gets a simple-function approximation of the elements $f_a$. In figure~\ref{fig:TRDpothomo}, 744 randomly generated tensor rank decompositions are taken and the points are ordered such that they represent a discrete version of the topology of the dual space. The minimal rank for the tensor rank decompositions here is $R=7$, thus the image represents $5208$ points of the circle. The resemblance with the $\sin(\theta)$ and $\cos(\theta)$ functions is clear, however if one actually zooms in one sees that the functions do deviate slightly. This is due to the randomness of the generation of tensor rank decompositions, and already signals that the metric one should find is not constant. This will be investigated more in section~\ref{sec:diffeo}. For more information on the procedure of generating potential homomorphisms through the tensor rank decomposition, and the reconstruction of the topology of the algebraic dual space, refer to~.

If a tensor $\tilde{P}_{abc}$ does not have a unit as described above, a unit may be constructed by using the eigen-problem of the tensor $\tilde{P}_{abc}$, meaning to find a solution to \footnote{This is a short summary of the procedure, for further discussion, see~\cite{Obster:2022oal}.}
\begin{equation}
    \alpha_a \alpha_b \tilde{P}_{abc} = \alpha_c.\label{eq:eigen}
\end{equation}
for some $\alpha_a \in \mathcal{F}$. This may be seen as a generalisation of the unit above. Contrary to the equation~\eqref{eq:def_unit}, a solution to~\eqref{eq:eigen} always does exist. Taking such a solution $\alpha_c$, one can construct a matrix
\begin{equation*}
    M_{ab} = \alpha_c \tilde{P}_{abc}.
\end{equation*}
Subsequently, one changes the basis such that this matrix is diagonal
\begin{equation*}
    M = diag(w_1, w_2, \ldots, w_N).
\end{equation*}
Here we assume that these $w_a$ are all positive. One can then define a new tensor (without Einstein-summation, in the basis where $M$ is diagonal)
\begin{equation}\label{eq:P_www}
    P_{abc} = \frac{1}{\sqrt{w_a w_b w_c}} \tilde{P}_{abc}, 
\end{equation}
which now has a unit given by $1=\sum_{a=1}^N \sqrt{w_a} \alpha_a f_a$.

It is precisely this procedure that gives a well-defined way of including spectral information of a Riemannian manifold into the tensor $P_{abc}$. Consider a positive compact operator $\mathcal{O}$ that depends on the Laplace-Beltrami operator, for instance
\begin{equation*}
    \mathcal{O} = \e^{-\Delta}.
\end{equation*}
If one would want to include the spectral data of this operator into the tensor $P_{abc}$ (which does have a unit), it could be done as
\begin{equation}
    \tilde{P}_{abc} = \braket{\mathcal{O}(f_c) |\, \mathcal{O}(f_a)\cdot \mathcal{O}(f_b)} = \e^{-\lambda_a} \e^{-\lambda_b} \e^{-\lambda_c} P_{abc},\label{eq:tildeP_laplace}
\end{equation}
where $\{\lambda_a\}$ are the eigenvalues of the Laplace-Beltrami operator. Given the tensor $\tilde{P}_{abc}$, one could then reconstruct both the tensor $P_{abc}$ (and thus the full topological structure) and the list of eigenvalues $\lambda_a$ by the procedure above and the identification
\begin{equation*}
    w_a = \e^{-2\lambda_a}.
\end{equation*}
Knowing these eigenvalues, we know the action of the Laplace-Beltrami operator on the basis functions. In the following section we assume the above-mentioned operator, such that our tensor $\tilde{P}_{abc}$ is assumed to be of the form~\eqref{eq:tildeP_laplace}. Note that the choice of this operator is a choice for a model. If one considers for instance the canonical tensor model, the ideal operator to interpret the results might differ slightly.

Lastly, let us comment on diffeomorphisms of the circle. Since the Laplace-Beltrami operator may be defined in a coordinate-free way~\cite{rosenberg_1997}, the eigenfunctions (up to diffeomorphisms) and eigenvalues are not expected to change. Furthermore, since the integration measure is also properly defined such that the integral is invariant under diffeomorphisms, a theory built from these tensors $P_{abc}$ is inherently coordinate independent. Therefore, the tensor of the ellipse is exactly the same as the tensor of the circle. 

\section{Reconstructing the metric from the Laplace-Beltrami operator}\label{sec:reconstruct}
In this section, it will be demonstrated how the metric of the circle or ellipses may be reconstructed. This is done using the the smooth functions on the circle and ellipse. For the circle, the metric can explicitly be calculated, but for the ellipse a numerical method will be developed. In section~\ref{sec:diffeo}, this method will then be applied to the algebraic tensor model approach.

First, let us discuss the case where the smooth functions of the circle are known. For the exact circle, the first few eigenfunctions are given in~\eqref{eq:circle_basis} with eigenvalues~\eqref{eq:eigen}. In order to calculate the metric at point $\theta_0$, $g(\theta_0)$, one would like to evaluate~\eqref{eq:circ_geom:operator} around a function $f_{\theta_0}(\theta)$ that overlaps around $\theta_0$ with
\begin{equation*}
    (\theta-\theta_0)^2.
\end{equation*}
In the case of the exact circle, one can verify that the function
\begin{equation*}
    f_{\theta_0}(\theta) = 2(1-\sin(\theta_0)\sin(\theta) - \cos(\theta_0)\cos(\theta))
\end{equation*}
around $\theta\sim\theta_0$ is given by
\begin{equation*}
    f_{\theta_0}(\theta) = (\theta-\theta_0)^2 + O((\theta-\theta_0)^4).
\end{equation*}
Knowing the eigenvalues of the Laplace-Beltrami operator, one can see that the action of the Laplace-Beltrami operator on this function is given by
\begin{equation*}
    \Delta f_{\theta_0}|_{\theta=\theta_0} = \sin(\theta_0)^2+\cos(\theta_0)^2 = 1,
\end{equation*}
which is indeed exactly the  metric given in~\eqref{eq:circ_geom:g}.

\begin{figure}
        \centering
        \begin{minipage}{0.475\textwidth}
            \includegraphics[width=0.95\textwidth]{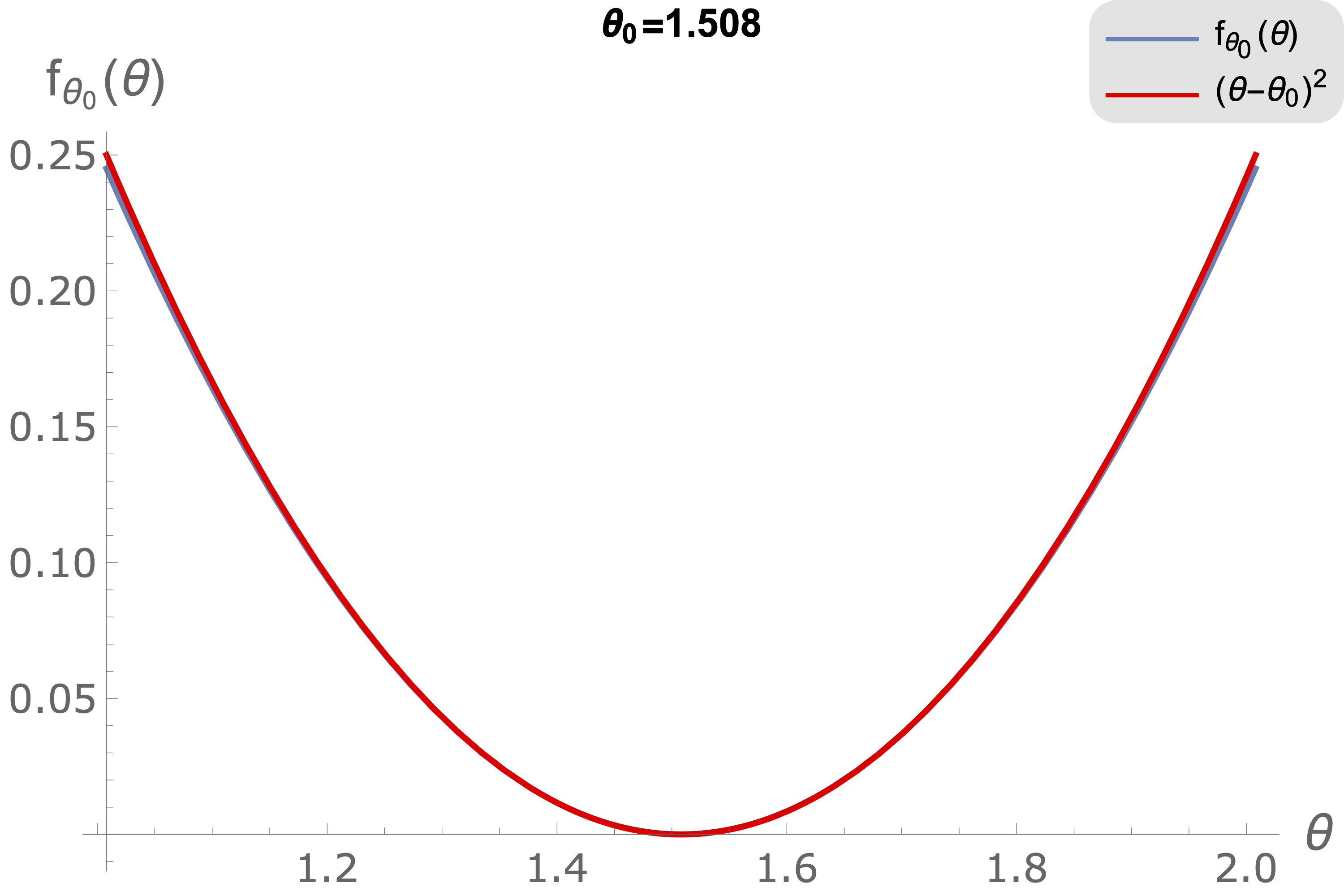}
        \end{minipage}
        \begin{minipage}{0.475\textwidth}
            \includegraphics[width=0.95\textwidth]{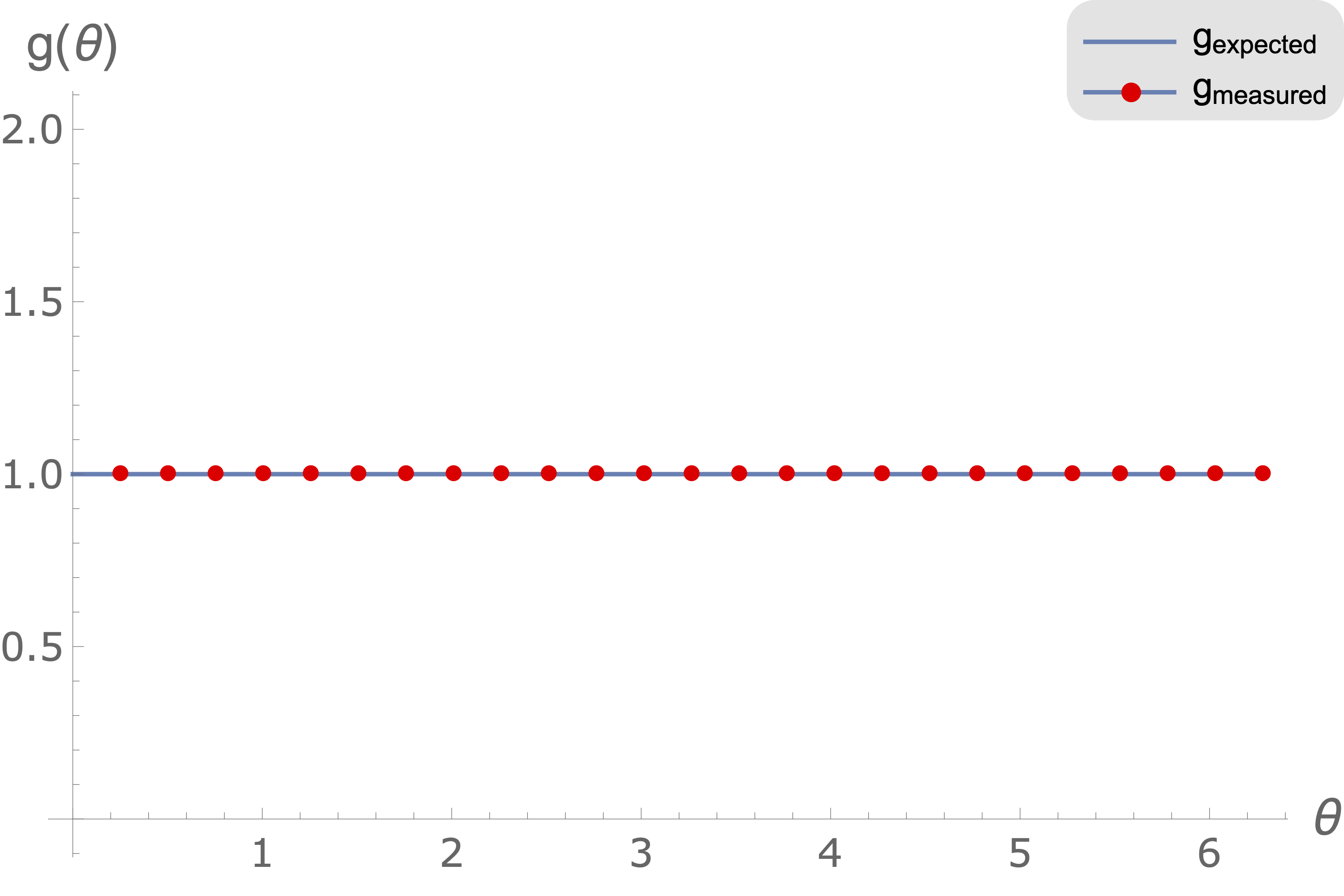}
        \end{minipage}
        \begin{minipage}{0.475\textwidth}
            \includegraphics[width=0.95\textwidth]{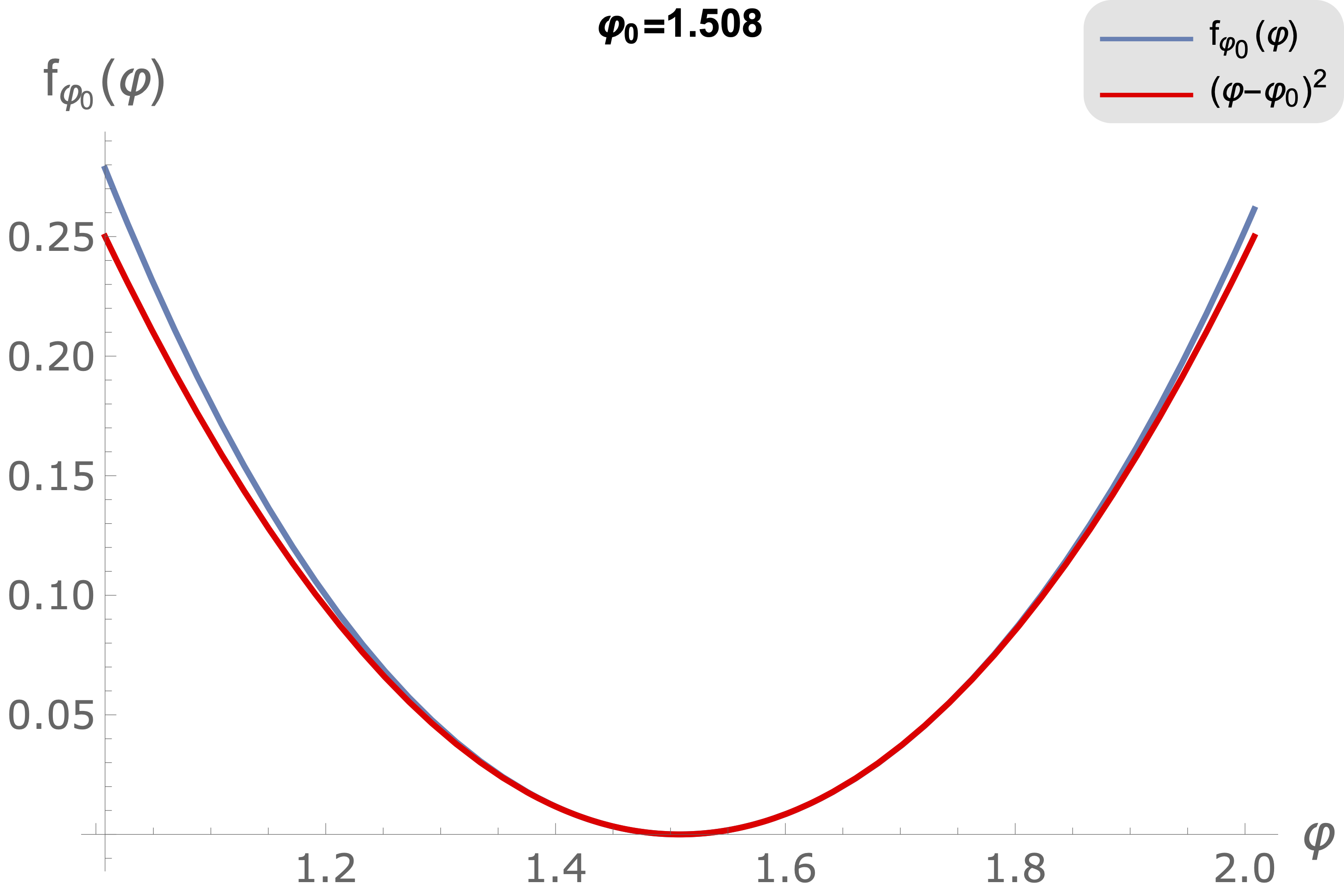}
        \end{minipage}
        \begin{minipage}{0.475\textwidth}
            \includegraphics[width=0.95\textwidth]{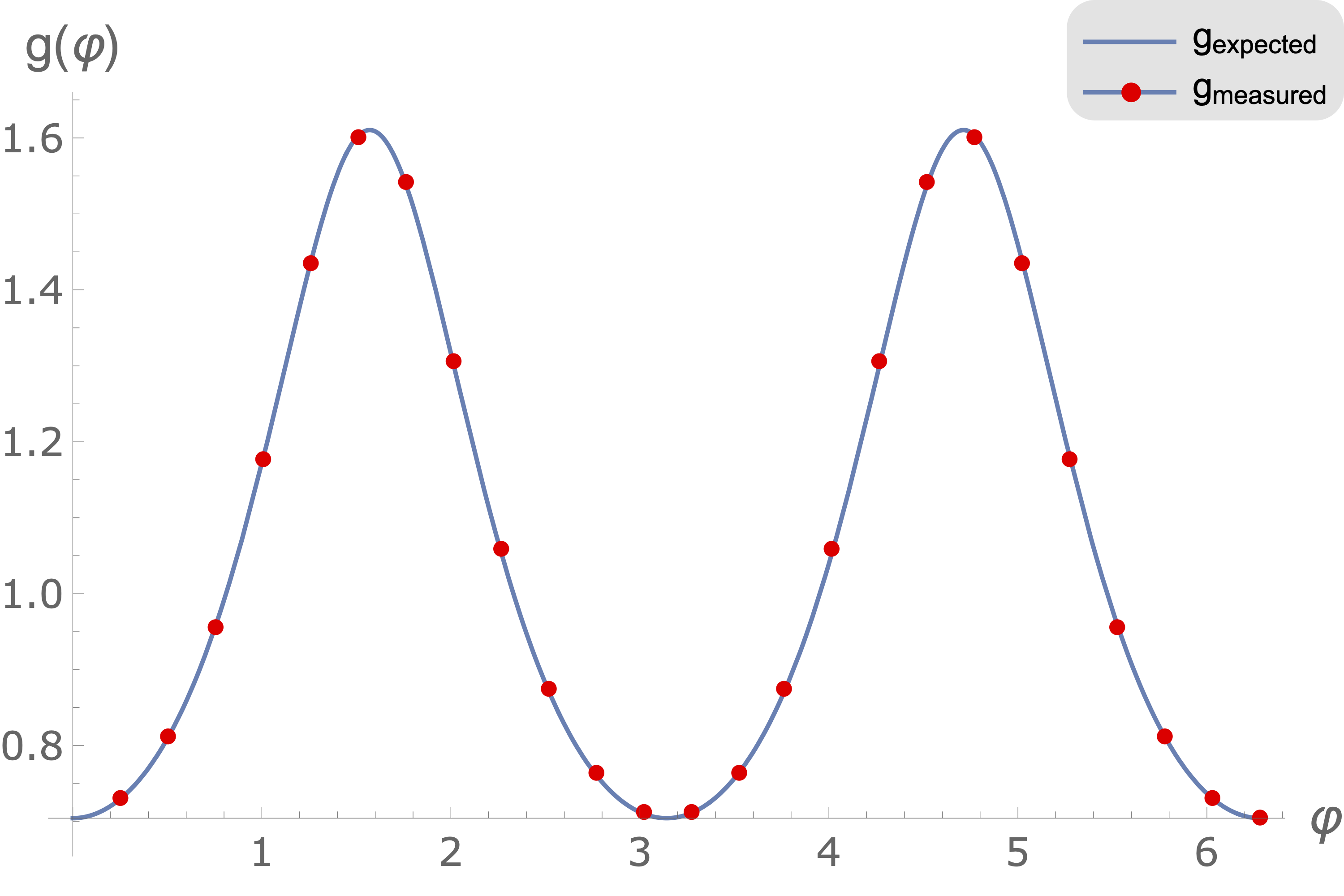}
        \end{minipage}
        
        \caption{The numerical calculation of the metric $g$ for the circle (top) and the ellipse (bottom). On the left, examples of the fitting functions $f_{\theta_0}$ and $f_{\phi_0}$ are shown in blue, that approximate $(\theta-\theta_0)^2$ (in red). On the right, the metric $g$ is plotted in both cases, with the measured values in red.}
        \label{fig:smooth_ellipse}
\end{figure}
One can do the same for the ellipse as defined in section~\ref{sec:exact_circle}. The new eigenfunctions are now given by\footnote{Note that the prime does \emph{not} denote the derivative.}
\begin{equation*}
    f_a'(\varphi) = f_a(\theta(\varphi)),
\end{equation*}
with $\theta(\varphi)$ as defined in~\eqref{eq:diffeo_ellipse}. Some examples of these functions are given in figure~\ref{fig:ellipse_example_functions}. One can now follow the same strategy, trying to  find an approximation of $(\varphi-\varphi_0)^2$ around a point $\varphi_0$. This is less straightforward as before, since there is no closed form of the incomplete elliptic integral of the second kind. However, numerically one can do this quite accurately. For this, take an $\epsilon>0$, and define the norm
\begin{equation}\label{def:norm:epsilon}
    \|f\|_{\varphi_0,\epsilon} = \int_{\varphi-\epsilon}^{\varphi+\epsilon} {\rm d}\varphi |f(\varphi)|^2. 
\end{equation}
For any point $\varphi_0$, one now wishes to find a function $f_{\varphi_0}$ that minimises
\begin{equation}\label{eq:norm_minimum}
    \| f_{\varphi_0}(\varphi) - (\varphi-\varphi_0)^2\|_{\varphi_0,\epsilon} .
\end{equation}
Then, one uses the action of the Laplacian on this function to get to know the metric at $\varphi_0$, as before. Note that the action of the Laplacian on these functions is the same as for the circle (given in~\eqref{eq:laplace_eigenv}), for instance
\begin{equation*}
    \Delta f_4 = 4f_4,
\end{equation*}
which does not depend on the coordinates one chooses. This (numerical) exercise was performed for $e=0$ (the circle) and $e=0.75$, and the result is shown in figure~\ref{fig:smooth_ellipse}. It can be seen that the reproduced metric is indeed exactly the metric of the circle and ellipse given in~\eqref{eq:circ_geom:g} and ~\eqref{eq:metric_ellipse}.

The metric of the exact circle is the easiest example imaginable, since it is simply constant. In the next section, diffeomorphisms of the circle will be considered, and it will be shown that it is possible to reconstruct the metric of an ellipse or randomly generated sets of points within the framework of the algebraic tensor models.

\section{The metric and diffeomorphisms in algebraic tensor models}\label{sec:diffeo}
In this section, a demonstration will be given of the calculation of the metric and the behaviour of diffeomorphisms in algebraic tensor models using the circle. First, the metric will be reconstructed for the circle and ellipse coordinate representations using the method developed in section~\ref{sec:reconstruct}. After this, an explicit diffeomorphism that smoothens out the functions resulting from a random collection of tensor rank decompositions will be taken.

\begin{figure}
        \centering
        \begin{minipage}{0.475\textwidth}
            \includegraphics[width=0.95\textwidth]{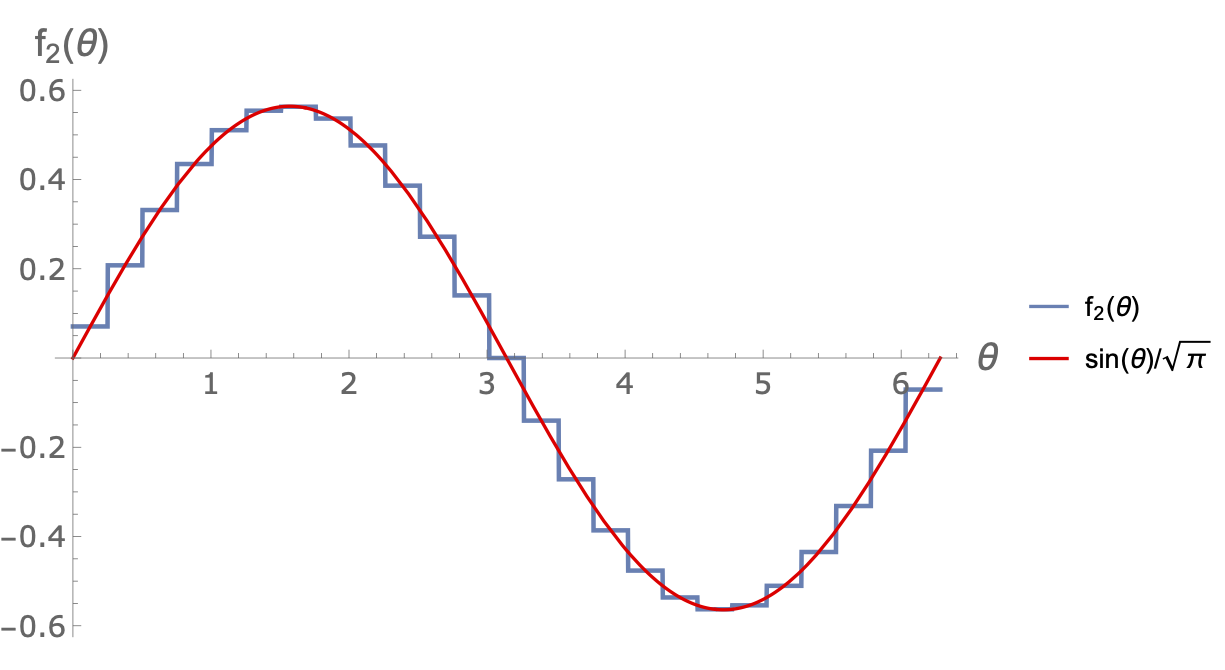}
        \end{minipage}
        \begin{minipage}{0.475\textwidth}
            \includegraphics[width=0.95\textwidth]{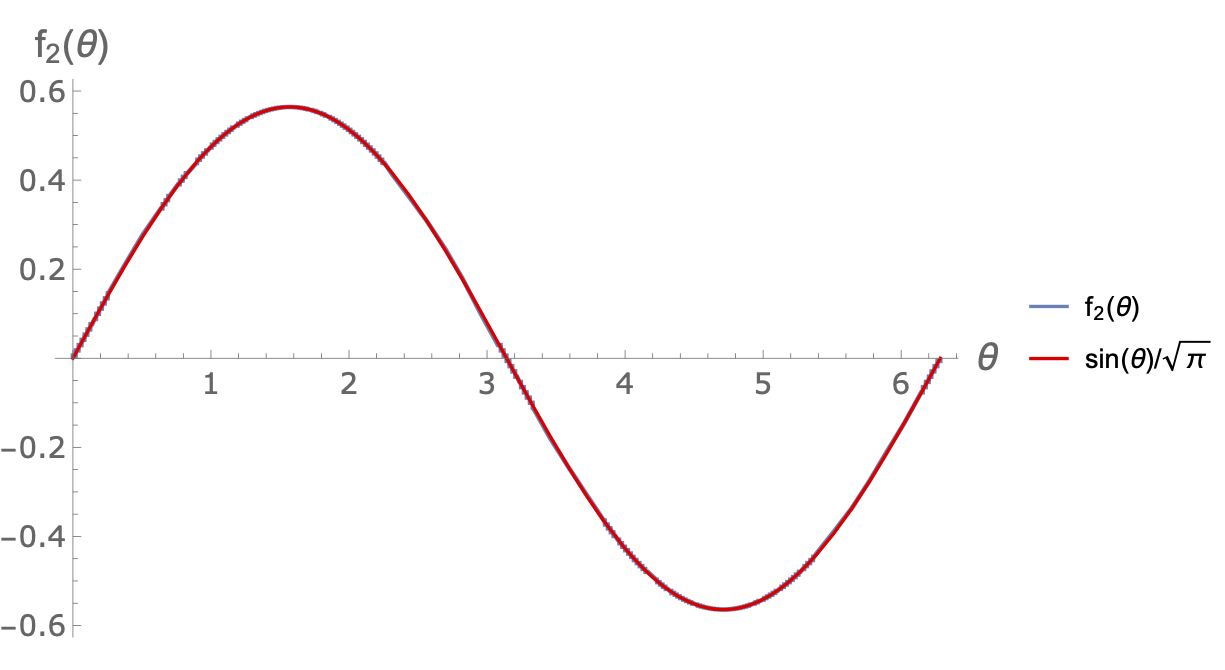}
        \end{minipage}
        
        \caption{Two examples of the simple function of~\eqref{eq:simple_f} for $R=25$ on the left and $R=300$ on the right. For large $R$ it can be seen that the function is a really good approximation of the function $\sin(\theta)/\sqrt{\pi}$.}
        \label{fig:perfect_circle_functs}
\end{figure}
Let us now assume we have a topological tensor $P_{abc}$ of the exact circle, and have already obtained a list of eigenvalues of the Laplace-Beltrami operator as described in section~\ref{sec:tensor_def}. To demonstrate the metric reconstruction in the algebraic tensor model approach, let us consider a tensor rank decomposition with perfectly spaced points
\begin{equation*}
    P_{abc} = \sum_{i=1}^R \beta_i p^i_a p^i_b p^i_c,
\end{equation*}
where
\begin{equation}\label{eq:p_beta_def}
\begin{aligned}
    \beta_i &= \frac{2 \pi\, i}{R},\\
    p_a^i &= f_a(\theta^i) = f_a(2\pi i/R).
\end{aligned}
\end{equation}
Here, the functions $f_a(\theta)$ are taken as in~\eqref{eq:circle_basis}. For $N=5$, the above tensor exactly reproduces the topological tensor defined in section~\ref{sec:tensor_def} if one takes $R\geq 7$. As explained in~\cite{Obster:2022oal}, it is useful to interpret the points $p_a^i$ as values of a simple function
\begin{equation}\label{eq:simple_f}
    f_a(\theta) = \begin{cases}
        p_a^1 \  , \ X_0 = 0\leq \theta < \beta_1\equiv X_1,\\
        p_a^2 \  , \ X_1\leq \theta < \beta_1 + \beta_2  \equiv X_2,\\
        \ldots\\
        p_a^R \  , \ X_{R-1} \leq \theta < \sum_{i=1}^{R} \beta_i \equiv X_R.\\
    \end{cases}
\end{equation}
Since this simple function will approximate the smooth functions better and better for larger $R$, if one wants to be able to approximate the function $(\theta - \theta_0)^2$ well, one needs a large amount of points $R$. In principle, one would like to take a continuous set to fully calculate the metric, but for practical reasons it suffices to take $R$ large enough. In figure~\ref{fig:perfect_circle_functs}, an example of these simple functions for $a=2$ is given, compared to its continuous counterparts. 

\begin{figure}
        \centering
        \begin{minipage}{0.475\textwidth}
            \includegraphics[width=0.95\textwidth]{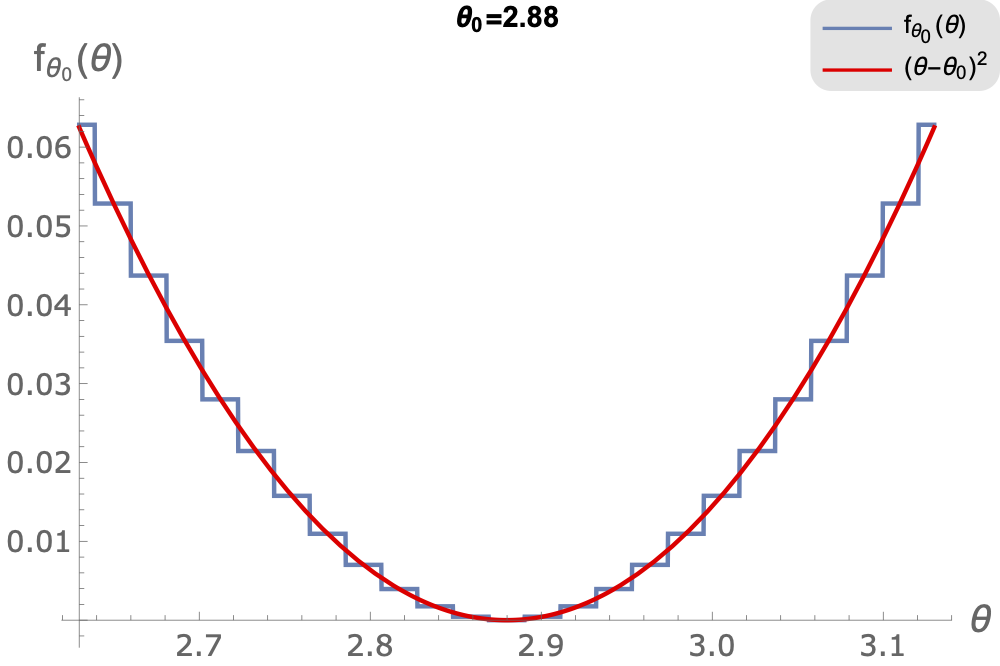}
        \end{minipage}
        \begin{minipage}{0.475\textwidth}
            \includegraphics[width=0.95\textwidth]{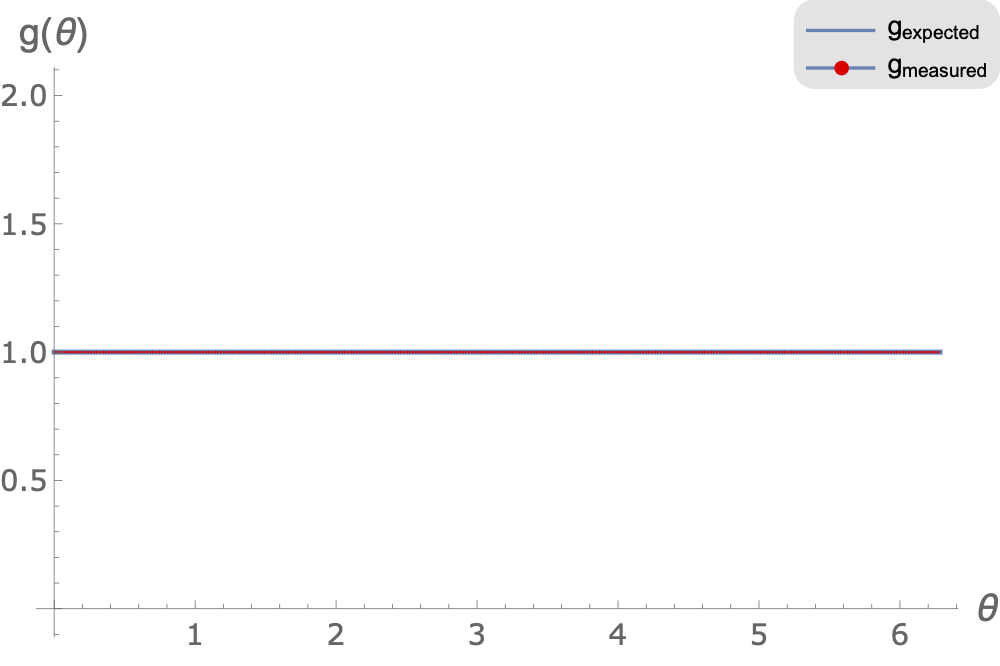}
        \end{minipage}
        \begin{minipage}{0.475\textwidth}
            \includegraphics[width=0.95\textwidth]{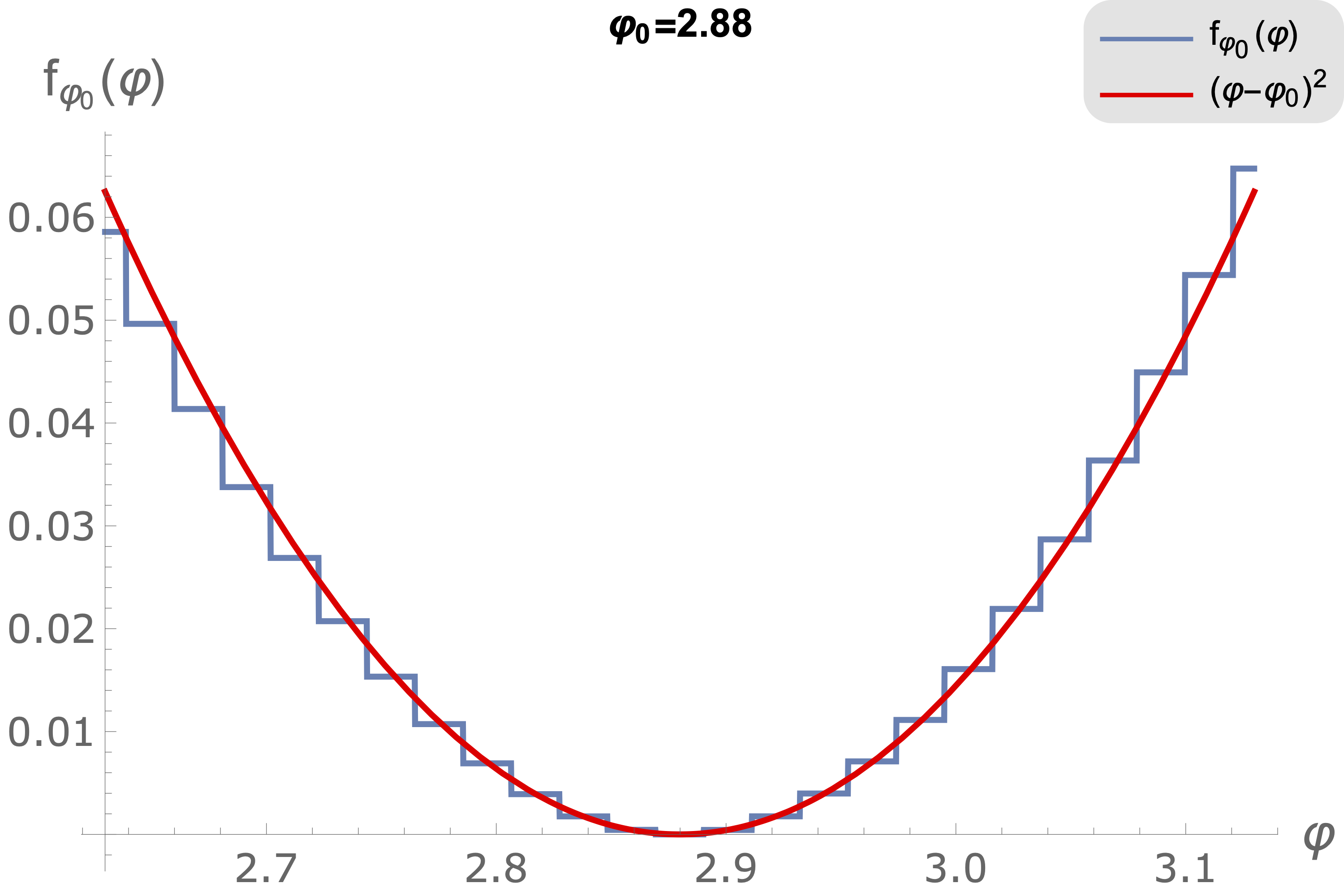}
        \end{minipage}
        \begin{minipage}{0.475\textwidth}
            \includegraphics[width=0.95\textwidth]{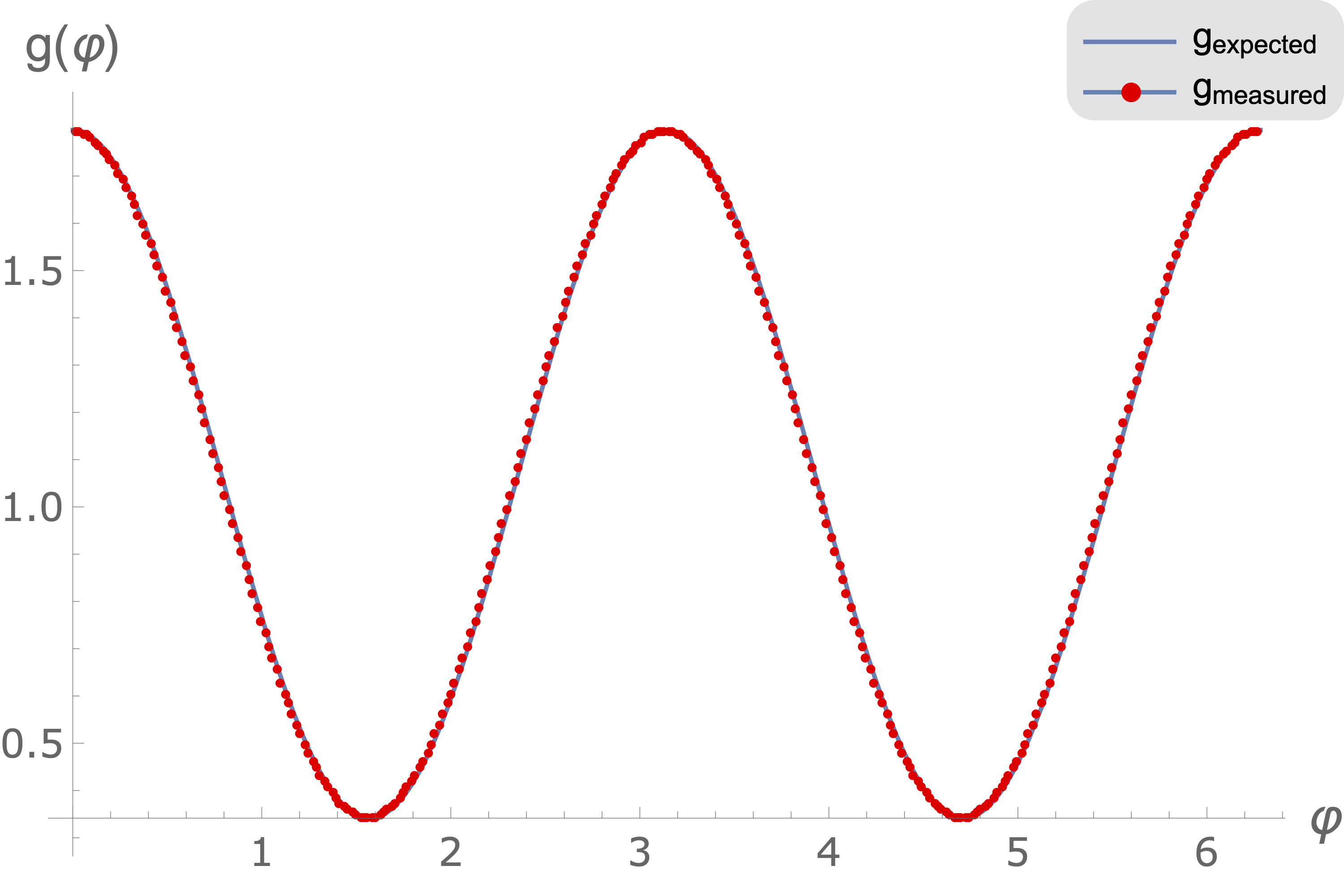}
        \end{minipage}
        
        \caption{The results of the calcualation of the metric of the circle (above) and the ellipse with parameter $e=0.9$ (below). On the left-side, an example fit function is shown that is used for the calculation of the metric at that point. On the right side, the calculated metric is given. In both cases, it is a perfect fit with the expected metric. The $\epsilon$-value of~\eqref{def:norm:epsilon} was taken to be 0.05.}
        \label{fig:perfect_ellipse}
\end{figure}

One can see that the integral in~\eqref{def:norm:epsilon} for the simple functions reduces to the sum
\begin{equation}\label{eq:norm_discrete}
    \|f\|_{\theta_0,\epsilon} = \sum_{i=1}^R B_{X_i,\theta_0,\epsilon} |f(X_i)|^2,
\end{equation}
where $X_i$ as in~\eqref{eq:simple_f} and 
\begin{equation*}
    B_{X_i,\theta_0,\epsilon} := \begin{cases}
        0 & \text{if } X_i > \theta_0+\epsilon \text{ or } X_i < \theta_0-\epsilon,\\
        \min(X_i,\theta_0+\epsilon) - \max(X_{i-1},\theta_0-\epsilon) & \text{otherwise}.
    \end{cases}
\end{equation*}
Note that for large $R$ this reduces to
\begin{equation*}
    B_{X_i,\theta_0,\epsilon} := \begin{cases}
        0 & \text{if } X_i > \theta_0+\epsilon \text{ or } X_i < \theta_0-\epsilon,\\
        \beta_i & \text{otherwise}.
    \end{cases}
\end{equation*}
One can again look for a function $f_{\theta_0} = \alpha^a_{(\theta_0)} f_a$ that minimises the norm~\eqref{eq:norm_minimum}, or in practice~\eqref{eq:norm_discrete}, and find the value of the metric by applying the Laplace-Beltrami operator  using~\eqref{eq:laplace_approx}. This was done using Mathematica, and the result of the metric is given in the upper part of figure~\ref{fig:perfect_ellipse}. The figure also shows an example of a function $f_{\theta_0}(\theta)$ that approximates $(\theta-\theta_0)^2$, to show the result of the fitting process. Note that the metric found is exactly the flat metric, as expected.

Furthermore, the ellipse will be considered. For this, one can consider a similar simple function as above, but by performing the diffeomorphism introduced in section~\ref{sec:exact_circle}, parameterised by $e$. To do this, one needs to redefine the simple function of~\eqref{eq:simple_f} according to the diffeomorphism~\eqref{eq:diffeo_ellipse}. This means that the simple function now becomes
\begin{equation}\label{eq:simple_f2}
    f_a(\varphi)' := f_a(\theta(\varphi)) = \begin{cases}
        q_a^1 \  , \ X_0 = 0\leq \varphi < \beta_1\equiv X_1,\\
        q_a^2 \  , \ X_1\leq \varphi < \beta_1 + \beta_2  \equiv X_2,\\
        \ldots\\
        q_a^R \  , \ X_{R-1} \leq \varphi < \sum_{i=1}^{R} \beta_i \equiv X_R,\\  
    \end{cases}
\end{equation}
with $\beta_i$ as in~\eqref{eq:p_beta_def} and $q_a^i$
\begin{equation*}
    q_a^i = f'_a(\varphi^i) = f_a(a(e) E(2\pi i/R,e)),
\end{equation*}
where $a(e)$ and $E(\varphi,e)$ were defined in section~\ref{sec:exact_circle}.\footnote{Note that one could also choose to implement the diffeomorphism by changing the $\beta_i$ in~\eqref{eq:simple_f} instead the values $q_a^i$.} Note that the tensor is diffeomorphism-invariant in the sense that 
\begin{equation*}
    P_{abc} = \sum_{i=1}^R \sqrt{g'_i} \beta_i q_a^i q_b^i q_c^i,
\end{equation*}
where
\begin{equation*}
    g'_i = g_{\varphi\varphi}(2\pi i/R),
\end{equation*}
with $g_{\varphi\varphi}$ as in~\eqref{eq:metric_ellipse}. One can now employ the exact same strategy as before, but now for the variable $\varphi$. This was again computed using Mathematica, and the result is given in the lower figures of~\ref{fig:perfect_ellipse}.

The above result shows how diffeomorphisms are to be regarded in algebraic tensor models. Since the models are formulated algebraically, there is no mention of coordinates yet. The models are therefore inherently diffeomorphism invariant. Diffeomorphisms can only be performed after one defined the coordinates of the functions by means of the algebraic dual space, as in~\eqref{eq:algebraic_dual}. The full geometry of the theory is purely encoded in the Laplace-Beltrami operator, which is defined in terms of the tensor in a model-dependent way as in~\eqref{eq:tildeP_laplace}.\footnote{Note that the operator under consideration in~\eqref{eq:tildeP_laplace} is $e^{-\Delta}$, but the exact choice of the operator depends on the model. The main important aspect is that one wants to fully reconstruct a list of eigenvalues of the Laplace-Beltrami operator from the eigenvalues of this operator.} 

An interesting note is that one could decompose the eigenfunctions of the ellipse in terms of the ``circle-functions'' $\sin(n\varphi)$ and $\cos(n\varphi)$, which corresponds to a Fourier-transform, for instance
\begin{equation}\label{eq:diffeo_as_basischange}
    \sin(\theta(\varphi)) = \sum_{n=1}^\infty a^n \sin(n \varphi).
\end{equation}
This is an interesting consideration to view the diffeomorphisms in this formalism in a different way, namely as a basis-change in the associative closure $\mathcal{A}\cong C^\infty(S^1)$. This has not been further investigated at this point, but it does imply that the diffeomorphisms $\text{Diff}(\mathcal{M})$ can be seen as a subset of the orthogonal transformations of the algebra of smooth functions $O(C^{\infty}(\mathcal{M}))$. It would be interesting to investigate this further in the future.

As a last example it will be shown that it is possible to ``smoothen out'' simple functions that were constructed from random tensor rank decompositions, a method discussed in section~\ref{sec:tensor_def}. Firstly one can find the metric $g_{\theta\theta}$ of this coordinate representation $\theta$. One can then use that the (smooth) metric transforms under diffeomorphisms as
\begin{equation*}
    g_{\varphi\varphi} = \frac{\partial \theta}{\partial \varphi} \frac{\partial \theta}{\partial \varphi} g_{\theta \theta}.
\end{equation*}
Therefore, if one wants the metric in the new coordinate system, $g_{\varphi\varphi}$, to be constant, one finds that the diffeomorphism has to satisfy
\begin{equation*}
\begin{aligned}
1 &= g_{\varphi\varphi} =  \frac{\partial \theta}{\partial \varphi} \frac{\partial \theta}{\partial \varphi} g_{\theta \theta}
&\Rightarrow \frac{\partial \theta}{\partial \varphi} = \frac{1}{\sqrt{g_{\theta \theta}}},
&\Rightarrow \varphi = \int \sqrt{g_{\theta\theta}} {\rm d}\theta .
\end{aligned} 
\end{equation*}
In the case of a discrete set of points $p_a^i$ with measure values $\beta_i$ as in~\eqref{eq:simple_f}, this effectively means that if one finds the discrete metric $g_i$ associated to every point with the procedure above, one could perform a discrete diffeomorphism
\begin{equation}\label{eq:discrete_diffeo}
    \beta_i \rightarrow \beta_i \sqrt{g_i},
\end{equation}
to arrive at a more smooth picture. 

One can do this for the example of the ellipse above, and indeed if one does this one gets back the eigenfunctions of the circle. More interesting to see what happens is the case of a random collection of points generated by tensor rank decompositions. As was found in figure~\ref{fig:TRDpothomo}, for a large number of tensor rank decompositions these functions look very similar to the circle. But if one zooms in one finds that there are some discrepencies coming from the fact that one randomly picks these tensor rank decompositions, instead of the perfect spacing between points as in~\eqref{eq:p_beta_def}. Note that ``random'' here is coming from the random starting conditions used to find tensor rank decompositions. Different distributions of these random starting conditions are expected to find different-looking functions that are related to each other through diffeomorphisms.

\begin{figure}
        \centering
        \makebox[\textwidth]{\makebox[1.2\textwidth]{\begin{minipage}{0.375\textwidth}
            \includegraphics[width=0.95\textwidth]{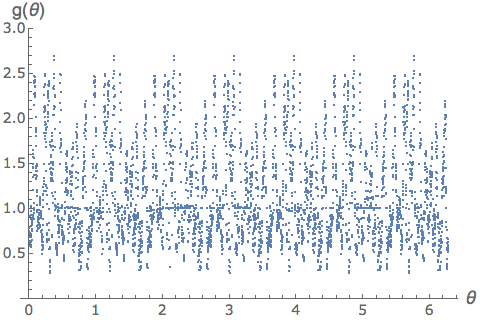}
        \end{minipage}
        \begin{minipage}{0.375\textwidth}
            \includegraphics[width=0.95\textwidth]{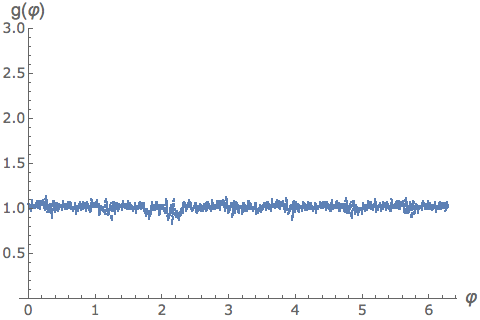}
        \end{minipage}
        \begin{minipage}{0.375\textwidth}
            \includegraphics[width=0.95\textwidth]{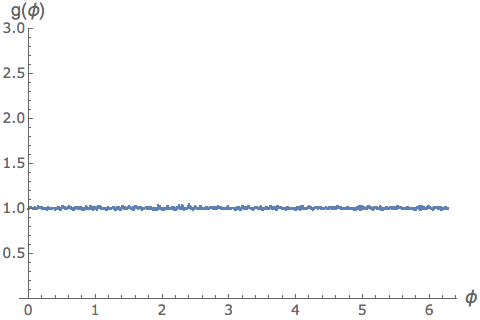}
        \end{minipage}}}
        
        \begin{minipage}{0.4875\textwidth}
            \includegraphics[width=0.975\textwidth]{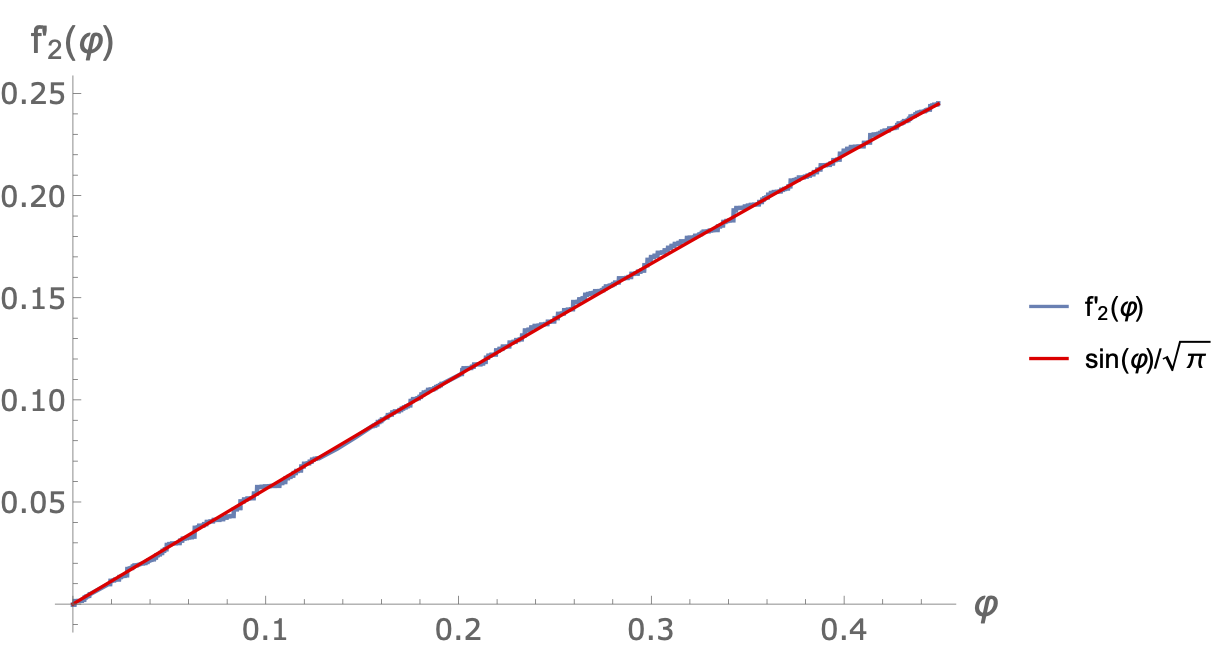}
        \end{minipage}
        \begin{minipage}{0.4875\textwidth}
            \includegraphics[width=0.975\textwidth]{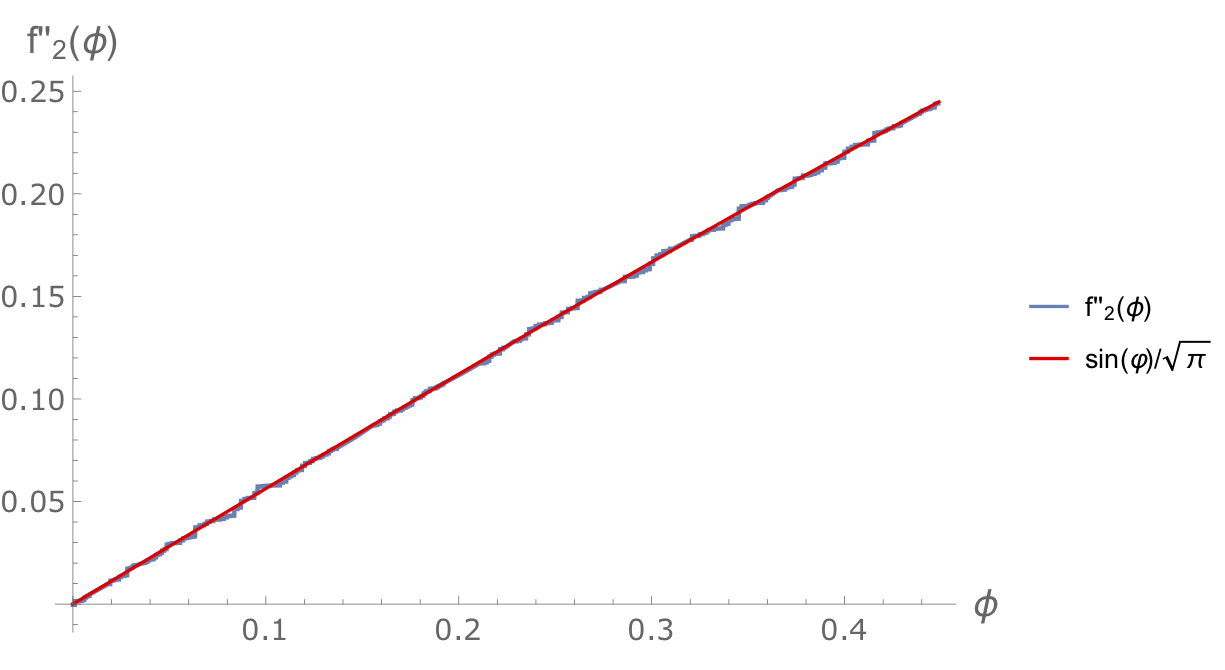}
        \end{minipage}
        
        \caption{The result of performing the discrete diffeomorphism as explained in the text. The upper three graphs show the metric. The left graph is in terms of the original coordinates in~\eqref{eq:simple_f} with $\epsilon=0.0125$, the middle graph in terms of new coordinates after one iteration with $\epsilon=0.05$, and the right graph after two iterations with $\epsilon=0.2$. The bottom half shows a comparison between the $\sin$-function and the functions $f'(\varphi)$ (one iteration) and $f''(\phi)$ (two iterations) respectively. Both are clearly more smooth than figure~\ref{fig:TRDpothomo}, and the second iteration seems even closer to the ideal $\sin$-function.}
        \label{fig:smoothen_out}
\end{figure}

An interesting aspect of this procedure now is the value of $\epsilon$. While before, this value was not strictly that important, it is expected to play a role here. This is because a small value of $\epsilon$ is expected to be able to detect smaller deformations, while large $\epsilon$ can be used to measure larger deformations. 

The procedure above done in order to smoothen out the functions in figure~\ref{fig:TRDpothomo}, and the results may be found in figure~\ref{fig:smoothen_out}. First, $\epsilon=0.0125$ was used to find the metric. It can be seen that this metric is still rather chaotic. However, if one takes a diffeomorphism as in~\eqref{eq:discrete_diffeo}, the function clearly becomes more smooth compared to figure~\ref{fig:TRDpothomo}. Subsequently, the metric was calculated using $\epsilon=0.05$. This metric already looks way more constant, as it deviates only slightly at every point. Lastly the metric was calculated using $\epsilon=0.2$, and in this case the metric looked rather smooth. It shows that, even with the random selection of tensor rank decompositions, it is possible to make a relatively smooth function in this way. Furthermore, the metric changes from having large peaks and being quite chaotic, to looking much more smooth.

\section{Summary}\label{sec:sum}
As found before in~\cite{Obster:2022oal}, these tensors can correspond to an associative algebra together with a list of eigenvalues of a self-adjoint compact operator, which is then used to reconstruct a list of eigenvalues of the Laplace-Beltrami operator. The (algebraic) dual space of this algebra, together with the Laplace-Beltrami operator was then argued to fully represent the geometric structure of a Riemannian manifold.

In this work, this was made explicit by fully investigating the case of the circle. It was shown that one can indeed reconstruct the metric as expected in an ideal scenario, where the points were evenly spaced. Furthermore, a class of diffeomorphisms, namely the ellipses, was considered, and it was shown that indeed the metric could be accurately reconstructed in these coordinate systems as well. Lastly, the metric was found for a random sampling of points that would be used in practice, where a random collection of tensor rank decompositions was used as introduced in~\cite{Obster:2022oal}. Moreover, it was shown that one can construct a diffeomorphism such that these points will be ``smoothened out''.

This work shows firstly that indeed the full geometric structure of a Riemannian manifold can be encoded in a tensor $P_{abc}$, and secondly it acts as a first investigation into the action of diffeomorphisms in this formalism. Furthermore, it explicitly shows how one can treat tensors relatively easily if they correspond exactly to Riemannian manifolds. It should be noted that more random tensors will be a bit more difficult, since one has to ensure that the elements $p_a^i$ in the tensor rank decomposition are indeed potential homomorphisms, as this has not been proven at this point.

There are quite some interesting research opportunities that follow. Firstly, it would be interesting to treat more examples than the - relatively simple - circle. For instance, one could repeat the analysis for the sphere or torus. A second interesting future consideration would be to further investigate the action of diffeomorphisms in this formalism as briefly mentioned around~\eqref{eq:diffeo_as_basischange}. Lastly it would be interesting to apply this knowledge to the canonical tensor model~\cite{Sasakura:2011sq}, by for instance using the equations of motion of the model and reconstructing the metric at every point in time, to see how the circle would evolve through time. Furthermore it would be interesting to know whether the circle appears in one of the peaks of the wave-function~\cite{Obster:2017pdq,Obster:2017dhx}.

\appendix

\section{The associative closure}\label{sec:app}
In this section, some notions introduced in~\cite{Obster:2022oal} that are used throughout the work are briefly reviewed. The main goal is to introduce the associative closure and potential homomorphisms. For a more extensive explanation, please refer to~\cite{Obster:2022oal}.

The first important concept is \emph{partial algebras}. Consider an algebra $(\mathcal{F},\cdot)$, a partial algebra is a sub-vectorspace $\mathcal{S}\subset \mathcal{F}$ with this same product. The crucial point here is that whereas $\mathcal{F}$ might not be an associative algebra, the partial algebra might actually be associative. Even if the full algebra $\mathcal{F}$ is non-associative, one can usually find an associative partial algebra. Partial algebras have an algebraic dual space, just like usual algebras, given by
\begin{equation*}
    |\mathcal{S}| := \{ p\in \mathcal{F}^*|\, p(f)p(g)=p(f\cdot g)\, \forall f,g \in \mathcal{S}\}.
\end{equation*}
Here, $\mathcal{F}^*$ is the linear dual space of $\mathcal{F}$. Note that $|\mathcal{S}|\subset\mathcal{F}^*$. The range of the partial algebra $\mathcal{K}^{(\mathcal{S})}$ is the sub-vectorspace that is covered by taking single products of the elements of $\mathcal{S}$. I.e.
\begin{equation*}
    \mathcal{K}^{(\mathcal{S})} := \{f\cdot g|\, \forall f,g \in \mathcal{S}\}.
\end{equation*}
A system of partial algebras $\{\mathcal{S}_i\}$ is a set of partial algebras such that for every two elements $S_i$ and $S_j$, the intersection of their algebraic dual spaces $|\mathcal{S}_i|\cap |\mathcal{S}_j| \neq \emptyset$. A system of partial algebras is called maximal if there is no system of partial algebras with a larger union of the ranges $\cup_i \mathcal{K}^{(\mathcal{S}_i)}$.

An associative extension is now defined as an algebra $\mathcal{A}$ consisting of a Hilbert space which is an extension of $\mathcal{F}$, i.e. $\mathcal{F}\subset \mathcal{A}$, and a product that is unital, commutative and associative, and reduces to $P_{abc}$ on $\mathcal{F}$ in the sense that
\begin{equation*}
    P_{abc} = \braket{f_c|\, f_a \cdot f_b},
\end{equation*}
with $f_a,f_b,f_c$ elements of a basis of $\mathcal{F}$. For consistency reasons, one also has to require that all of the elements of the dual space $p\in |\mathcal{A}|$, when projected to $\mathcal{F}^*$, are included in the dual space of one of the maximal systems of partial algebras.

A \emph{potential homomorphism} is now defined as the projection of an element of the dual space of some associative extension, $p \in |\mathcal{A}|$, to the linear dual space $\mathcal{F}^*$, i.e. $p^* = p|_{\mathcal{F}^*}$. The collection of all potential homomorphisms (of all possible associative extensions) is the \emph{space of potential homomorphisms}.

Finally, the \emph{associative closure} is the associative extension whose dual space is isomorphic to the space of potential homomorphisms. This is in that sense the maximal associative extension possible.

In~\cite{Obster:2022oal} a conjecture was stated that for algebras that represent spaces of functions over some compact manifold, the tensor rank decomposition can be used to find potential homomorphisms
\begin{equation*}
    P_{abc} = \sum_{i=1}^R \beta_i p_a^i p_b^i p_c^i,
\end{equation*}
as long as the tensor rank decomposition is taken to be \emph{positive} and \emph{minimal}, meaning that all $\beta_i>0$ and $R$ is the lowest natural number possible. For any tensor rank decomposition, $p_a^i$ can be interpreted as elements of $\mathcal{F}^*$, and the conjecture states that in this case $p_a^i$ are all potential homomorphisms. An explicit recipe for reconstructing the associative closure when one has the space of potential homomorphisms is given in~\cite{Obster:2022oal}.

\printbibliography

\end{document}